\documentclass{Interspeech}

\interspeechcameraready 

\usepackage{hyperref}

\title{Benchmarking Time-localized Explanations for Audio Classification Models}

\author[affiliation={1,2}]{Cecilia}{Bolaños}
\author[affiliation={1,2}]{Leonardo}{Pepino}
\author[affiliation={1,2}]{Martin}{Meza}
\author[affiliation={2}]{Luciana}{Ferrer}
\affiliation{Departamento de Computación, FCEyN}{UBA}{Argentina}
\affiliation{Instituto de Investigación en Ciencias de la Computación (ICC)}{CONICET-UBA}{Argentina}
\email{cbolanos@dc.uba.ar, lpepino@dc.uba.ar, mmeza@dc.uba.ar, lferrer@dc.uba.ar}

\keywords{explainability, model-agnostic post-hoc explanations, benchmark, spurious correlations, audio models}

\usepackage{comment}
\begin{document}

\maketitle
\begin{abstract}
    Most modern approaches for audio processing are opaque, in the sense that they do not provide an explanation for their decisions. For this reason, various methods have been proposed to explain the outputs generated by these models. Good explanations can result in interesting insights about the data or the model, as well as increase trust in the system. Unfortunately, evaluating the quality of explanations is far from trivial since, for most tasks, there is no clear ground truth explanation to use as reference. In this work, we propose a benchmark for time-localized explanations for audio classification models that uses time annotations of target events as a proxy for ground truth explanations. We use this benchmark to systematically optimize and compare various approaches for model-agnostic post-hoc explanation, obtaining, in some cases, close to perfect explanations. Finally, we illustrate the utility of the explanations for uncovering spurious correlations.
\end{abstract}
\section{Introduction}

Explaining a machine learning model's decisions can help improve the users' trust in the system by providing an understanding of the patterns that led to them \cite{nasim2024trustworthyxaiapplication}. Further, explanations may provide interesting insights about the task or uncover hidden biases or spurious correlations \cite{adebayo2020debugging}. Explainability is particularly important in high-stakes applications such as disease detection, speaker verification, or emotion recognition~\cite{islam2022study,Ben-amor2023,nfissi2023iterative}. 
In machine learning, post-hoc explainability methods aim to interpret the output of opaque models after they are trained. These techniques are categorized into model-specific and model-agnostic methods \cite{surveyxai}. Model-specific methods leverage internal details of the model’s architecture, such as gradients in deep neural networks or attention mechanisms in transformers \cite{LRP,deeplift,IG}. These methods are tightly coupled with the model's design and cannot be applied to other types of models.  In contrast, model-agnostic methods do not assume any specific architectural properties of the model and treat it as a black box, relying solely on input-output interactions \cite{ribeiro2016should, scott2017unified, CFE}. 
In audio, black-box explainability methods rely on the adaptation of general frameworks like LIME \cite{ribeiro2016should} and KernelSHAP~\cite{scott2017unified}, using a common approach. First, perturbations of the input sample are generated and the model is run on each of them. Then, the perturbation patterns and the model outputs are used to calculate the importance of specific features or segments in the signal. For instance, SLIME \cite{mishra2017local} applies LIME on audio tasks by dividing the signal into temporal, spectral, or time-frequency blocks for perturbation.  AudioLIME \cite{haunschmid2020audiolime} extends this approach by incorporating source separation to decompose a music track into stems like vocals and drums before perturbation. Finally, 
TimeSHAP \cite{bento2021timeshap} builds upon KernelSHAP to handle sequential data. 
Evaluating explanation approaches is a difficult task since it is generally not possible to obtain ground truth explanations. In the works described above, evaluation is done through qualitative inspection \cite{mishra2017local,bento2021timeshap} or by measuring the impact of the most important components identified by the explanation on the model's output. The latter approach uses a set of metrics, often referred to as faithfulness (FF), which assess the effect of removing or incorporating these components \cite{chan-etal-2022-comparative,paissan2024listenablemapsaudioclassifiers, parekh2022listeninterpretposthocinterpretability, Akman2024AttHearEA}. FF metrics, though, have been shown to be poor assessors of explanation quality since they can reward explanations that rely on suppressor variables rather than informative ones \cite{haufe2024explainableaineedsformal, wilming2023theoreticalbehaviorxaimethods}. 

In this work, we study methods that provide time-localized explanations for black-box audio models and propose a benchmark for evaluating the explanations consisting of tasks where the goal is to detect or count the occurrences of a particular event. We use real and synthetic datasets where the location of the events of interest is known. These annotations are used as a proxy ground truth for the explanations, assuming that the important segments should be strictly contained within the regions where the target events occur.  We use this benchmark to optimize and compare various approaches. Furthermore, we illustrate the practical utility of these methods to uncover spurious correlations in audio datasets. 

\begin{figure}[t]
\centering
\includegraphics[width=0.97\columnwidth]{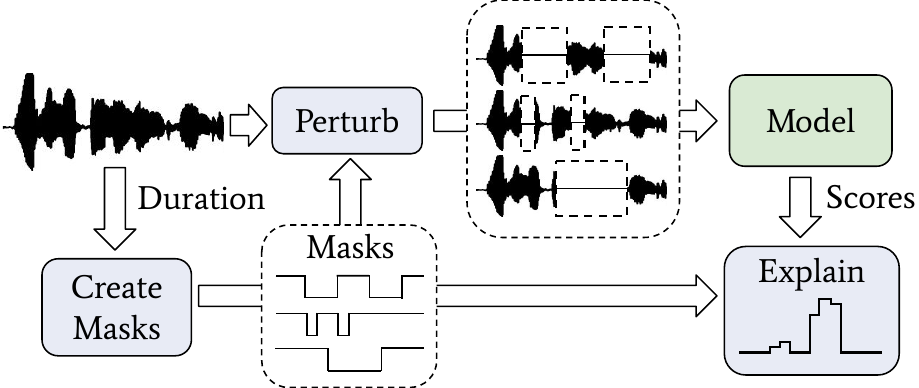}
\vspace{-3mm}
\caption{Procedure to produce explanations for a given black box model and input signal. First, a set of masks are created. Then, each mask is used to create a perturbed version of the input signal, which is fed into the model. The resulting scores, together with the corresponding masks, are used to estimate the impact of masking each segment of the input signal.}
\vspace{-6.5mm}
\label{fig:pipeline}
\end{figure}

\section{Methods}
\label{sec:methods}

In this study, we evaluate black-box explainability methods for audio models. In particular, we focus on time-level explanations, i.e., those capable of identifying the segments within an audio clip that explain a model’s prediction, as opposed to feature-level explanations that aim to identify the audio characteristics that are important for the model.  To evaluate the explanations, we would like to have a ground truth annotation for each signal that identifies the parts of that signal that are important for prediction. Unfortunately, such ground truth is often hard to annotate. 
While humans are able to provide labels for most audio classification tasks with varying degrees of agreement depending on the task, it may be much harder to reliably pinpoint the exact temporal regions that explain their decision. To complicate matters more, the ground truth may be model-dependent since different models may focus on different regions of a signal to make their decisions.

To work around the need to annotate ground truth explanations, in this work we propose to benchmark explanations on event detection and counting tasks where the location of the events of interest is known. We then use those locations as a proxy for the ground truth explanation, assuming that the important information for the models is \emph{only} present within those regions and that no part of them is unimportant. While the first assumption is reasonable for our selected tasks, the second one is unlikely to hold. As we will see, though, our results suggest that an inaccuracy of this kind in the ground truth does not change the ranking of systems, validating the value of the metric for development purposes.

\subsection{Time-localized explanations for audio models}

A general strategy for producing temporal explanations for black-box audio models consists of three basic steps: (1) creating several perturbed versions of the sample for which the explanation is required, by masking chunks of the audio, (2) running the black-box model on each perturbed sample to generate scores, and (3) training a surrogate model on the perturbations and their corresponding scores to estimate the importance of each segment. This process is illustrated in Figure~\ref{fig:pipeline}. The computation requirements for obtaining an explanation for a given sample and model are approximately the same as running inference $N$ times, where $N$ is the number of perturbed versions of the sample we use to train the surrogate model. In general, the training of the surrogate model takes a negligible amount of time.



In our implementation, the audio signals are first segmented into 100 ms segments 
and the signals are perturbed as follows:
\begin{enumerate}
    \item Initialize a binary mask vector \( \mathbf{v} \in \{0, 1\}^M \), where $M$ is the number of segments in the signal, with all ones.
\item While the number of masked segments (the number of 0s in $\mathbf{v} $) is less than \( T = \lceil M \times p \rceil\), where $p$ is the fraction of the waveform to perturb:
    \begin{enumerate}[topsep=0pt,itemsep=-1ex,partopsep=1ex,parsep=1ex,label=\alph*.]
        \item Select a random starting position \( s_i \) between 0 and $M$.
        \item Mask a contiguous block of \( d \) segments starting at $s_i$: \(\mathbf{v}[s_i : \min(s_i + d, M)] \leftarrow 0.\)
        \item Exclude $s_i$ from future selection as starting position.
    \end{enumerate}

    \item Replace all segments where the mask $\mathbf{v}$ is set to 0 with zeros or with gaussian noise  with mean equal to the mean of the original signal and a standard deviation sampled uniformly between 0.1 and 1.0 times the average standard deviation of the signals in the dataset.
\item Finally, ground truth labels are assigned to each segment.  If the overlap between the segment and the annotated region is larger than 90 ms, that segment is labeled with a 1, if there is no overlap, it is labeled with a 0, otherwise, the segment is ignored for performance evaluation. 
\end{enumerate}

Given an audio signal for which the explanation is required, a set of $N$ perturbed signals is created with this algorithm, varying the $p$ and $d$ values. Each of these perturbed signals is run through the black-box model \( f\) and a score is computed as the log-odds 
$\log(P_c/(1-P_c))$, where $P_c$ is the posterior probability produced by the model for class $c$, the class detected by the model when fed with the original signal.
The final product of this process is a dataset consisting of $N$ mask vectors $\mathbf{v}$ and corresponding scores (see Figure \ref{fig:pipeline}). 

Finally, a surrogate model \( g\) is trained using this data to minimize the discrepancy between its outputs and the scores obtained from the black-box model, taking the mask vectors as input features. This is achieved by solving the following optimization problem \cite{ribeiro2016should}:
\begin{equation}
\xi = \arg\min_{g \in \mathcal{G}} L(f, g, \pi) + \Omega(g),
\end{equation}
where $\mathcal{G}$ is the space of surrogate models under consideration, \( \Omega(g) \) is a regularization term, and 
\( \pi \) is a weighting function.  

Variations in the choice of $L$, $\mathcal{G}$, \( \Omega \), \( \pi \) and the way to perturb an instance give rise to different explainers found in the literature. When \( g \) is a linear model, \( \Omega \) is some norm of that model's weights, \( \pi \) is a kernel function applied over the distance between the perturbed and the original samples, the loss is weighted mean squared error, and the masking is done with zeros, the method becomes a variant of LIME, as used in~\cite{mishra2017local, haunschmid2020audiolime}.   When, instead, \( g \) is a Random Forest and the masking is done with noise, the approach resembles that in \cite{abdullah2024sig}. We refer to these approaches as \textbf{LR} and \textbf{RF} and explore both types of masks for both modeling approaches. 

When, instead, \( \pi \) is obtained as function of the number of zeros in $\mathbf{v}$, and  $\mathcal{G}$ is restricted to linear models where the intercept must be \( f(\emptyset) \) (the model's output when $\mathbf{v}$ is all zeros) and the sum of the weights must equal the model's score for the original signal, the method is called Kernel SHAP \cite{scott2017unified}. We refer to this approach as \textbf{SHAP}. This method was applied to audio signals in \cite{bento2021timeshap} where they use a structured perturbation approach, replacing feature values with their average values. 

The final step to produce the explanations is to use the surrogate model to estimate the importance of each segment in the signal. The way to obtain the importance of each feature and, hence, each segment, depends on the choice of $\mathcal{G}$. For linear models as LR and SHAP, the weights directly reflect the importance of each feature. For RF, the importances can be estimated as the normalized total reduction of the Gini criterion brought by each feature. 
In this work, the quality of these explanation for each signal is assessed using the area under the ROC curve (AUC) with respect to the labels obtained as described above. The final performance metric is obtained as the average AUC across signals. 
\subsection{Benchmark Datasets and Models}

In this section we describe the datasets and classification models for which explanations will be produced and evaluated. The main guideline for selecting datasets is that they should allow for an audio classification task and that the location of the events corresponding to the classes should be available. All the datasets described below are available in the paper's website.\footnote{\href{https://sites.google.com/view/time-localizedexplanations/}{https://sites.google.com/view/time-localizedexplanations/}}

\label{sec:kws}
The {\bf KWS} dataset consists of samples extracted from the Librispeech dataset \cite{panayotov2015librispeech}. The classification task is to detect the word "little", one of the most frequent content words in this dataset. As the black-box classification model, we trained a deep neural network using features extracted from wav2vec 2.0 base. Activations from the wav2vec 2.0 CNN encoder output and the 12 transformer blocks are extracted and combined by learning a weighted average. Finally, a mean pooling over time is performed and the resulting vector serves as input to a multilayer perceptron (MLP) with 128 hidden neuron, which predicts the probability that the word ``little'' is present in the audio. The model is trained with 500 utterances that include the target word and 500 utterances that do not include it, extracted from the 100 hours clean training set of Librispeech. We used all the utterances that include the target word in the dev-clean and test-clean sets, and the same number of utterances that did not include this word, to select the best epoch and test the model, respectively. The trained model achieved a 82\% accuracy in this keyword spotting task. The ground truth annotations were obtained with the Montreal Forced Aligner \cite{lugosch2019speechmodelpretrainingendtoend, mcauliffe17_interspeech}.

We also use a subset of the AudioSet Temporally-Strong Labels dataset \cite{hershey2021benefit}, which provides precise temporal annotations of sound events, performed by human annotators. From this data, we select three frequent events: {\bf Speech}, {\bf Music}, and {\bf Dog} sounds. As detection model, we use the Audio Spectrogram Transformer (AST) \cite{gong21b_interspeech}, a transformer-based model fine-tuned on AudioSet, which has a binary output for each of 527 possible events. We only use the outputs corresponding to the three selected event types.

Further, we generated a synthetic dataset by concatenating {\bf Drum} sounds obtained from \cite{drum_dataset}. We randomly generated 6000 sequences of 10 drum hits from 4 different types: kick, snare, toms or cymbals. Samples contain between 0 and 5 kicks, including 1000 samples per case. The placement of the kicks within each sequence is randomly selected and the chosen positions are used to determine the ground truth annotations. The model is similar to the one used for KWS, but an LSTM with 256 units is added before the MLP. The model is trained to predict the number of kicks in the waveform, using 4000 samples as training set, and 1000 samples for validation and testing. The resulting model achieves a 97\% accuracy.

The drum hits that were concatenated to create the audio samples for the Drums dataset contain most of the energy in the onset, when the instrument is hit, and then the energy decays fast with time. Hence, the ground truth annotations for the kick locations include regions that are almost silent and may be irrelevant for classification.
A similar observation can be made for audioset annotations, though in that case we cannot reliably assume that the important part is always at the beginning of the annotation.
In Section~\ref{sec:impact-inaccurate-gt}, we analyze the impact of this issue on the explanation metric.

\subsection{An application: Clever-Hans effect detection} \label{sec:cleverhans}

Clever-Hans effect, in the context of machine learning, occurs when a model's decisions are based on information that correlates in a spurious way with the actual information of interest. These kinds of spurious correlations have been shown to occur in real-world datasets \cite{liu24f_interspeech, arjovsky2020invariantriskminimization,borah2024measuringspuriouscorrelationclassification, WALLIS2022102368, gauder2024unreliabilityacousticsystemsalzheimers}. Explanation approaches can help pinpoint such issues. To test whether the models developed in this work can be used for this purpose, we generated a modified version of the IEMOCAP dataset \cite{busso2008iemocap} by adding coughs only to the utterances belonging to the ``happy'' class. A model trained on this corrupted dataset will likely rely on detecting the presence of cough to predict the ``happy'' class.  
For the utterances annotated as ``happy'', we randomly sampled one of 40 coughs from the ESC50 dataset \cite{piczak2015esc}, normalized them and deleted the silences based on energy levels, before summing them in a random location of an IEMOCAP utterance. 

We trained our model using sessions 1 through 4 and used the fifth session as test set. We did not perform hyperparameter tuning so no validation set was needed. Our model is the same as the one used for KWS except that the output layer contains 4 output nodes, one per emotion: happy, angry, sad, and neutral. Training with the original signals yielded an AUC of 0.761 for ``happy'', which rose to 0.943 after adding coughs—inducing a Clever-Hans effect we aim to reveal with an explainer model.


\section{Results and Discussion}

In this section we first show results for the various tasks in our dataset comparing  different explainability approaches. We then explore the effect of inaccurate ground truth annotations and illustrate the use of explainability models for the detection of spurious correlations.

\subsection{Benchmarking explainability approaches}

\begin{figure}[t]
\centering
\includegraphics[width=\columnwidth]{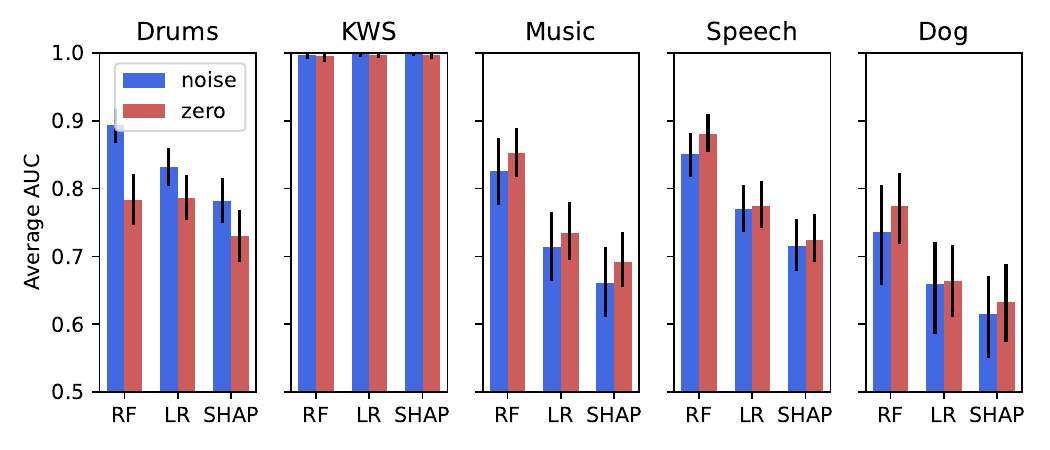}
\vspace{-8mm}
\caption{Average AUC for the datasets in our benchmark for different explanation approaches. The whiskers correspond to the 95\% confidence interval obtained by bootstrapping the test samples~\cite{ferrer2024goodpracticesevaluationmachine}.}
\label{fig:auc_relaxed}
\vspace{-6.5mm}
\end{figure}

Figure \ref{fig:auc_relaxed} shows the results for three surrogate models, RF, LR and SHAP as described in Section \ref{sec:methods}. In our preliminary experiments, we found that the kernel weighting function and the regularization term  did not have an impact in the LR results, so here we only show results for standard linear regression. Further, for SHAP we found that the constraint that the bias term should coincide with the model's output for a fully-masked signal significantly degraded results. Hence, in our runs, we eliminated this constraint. For each type of surrogate model, we show results using zeros or gaussian noise for masking.

The perturbed signals are obtained by setting  $d$, the size of each masking window in segments, to 1, 3, or 5; and a perturbation proportion $p$ to 0.2, 0.3, or 0.4. Detailed results for each combination of these parameters can be found in the \href{https://sites.google.com/view/time-localizedexplanations/}{paper's website}. Here, we show results when training the surrogate models using a total of 3000 samples with approximately equal number for each combination of $p$ and $d$.

The results in Figure \ref{fig:auc_relaxed} are obtained only on samples that contain the events of interest and the model correctly predicts the class. For the Audioset datasets, since the task is harder than for KWS and Drums, we keep only samples that had a score for the target class in the top 50\%. We explain only correctly predicted samples, as, for other cases, the model may use information beyond the ground truth, invalidating the evaluation procedure. The final sets used to obtain results include 50, 50, 52, 71, and 44 samples for Drums, KWS, Music, Speech, and Dog, respectively.

The figure shows that all explainers reach an average AUC well above random performance (0.5 for AUC). The best surrogate model, in all cases, is RF, suggesting that linear models are not able to adequately model the interactions between different segments of the signal. Notably, the optimal type of mask depends on the dataset with noise being better for Drums and zeros being better for all Audioset sets. The similarity between results on the three Audioset sets is likely explained by the fact that the same model, AST, is being explained in all cases. This suggest that the optimal type of mask may depend on the model to be explained, yet more experimentation is needed to support this hypothesis.

A notable result in Figure \ref{fig:auc_relaxed} is that the average AUC on the KWS dataset is almost 1.0 for every explainer. This result validates our benchmarking procedure since it shows that obtaining perfect average AUC with an explainability model is possible. Further, it indicates that the model needs to rely on the full word to make its decision in order to differentiate it from all other words. For other tasks, models may rely on only part of the event, but our evaluation compares explanations to the entire segment, leading to lower AUC even when explanations are accurate. In addition, the perfect result in KWS suggests that explanations for models that were trained to detect events without time-aligned annotations can be used as time-aligners for those events. This is a very interesting conclusion given that temporal annotations of acoustic events are expensive to obtain. 

The fact that our evaluation approach requires datasets with time annotations significantly restricts the datasets that can be used for benchmarking. The faithfulness (FF) metrics, on the other hand, can be computed in an unsupervised way. If such metric correlated well with the average AUC we could use it as a proxy, avoiding the need for time annotations. Unfortunately, this is not the case. In our experiments, the average FF computed as the difference between the score obtained with the unperturbed signal and the one obtained after masking the top $X\%$ most important segments \cite{chan-etal-2022-comparative} has very different trends compared to the average AUC for any value of $X \in \{1, 5, 10, 20, 50\}$. For example, according to FF, SHAP is the best method for Drums, while, according to the average AUC, SHAP is the worst method. 
Detailed results on FF can be found in the paper's website.

\begin{figure}[t]
\centering
\includegraphics[width=0.9\columnwidth]{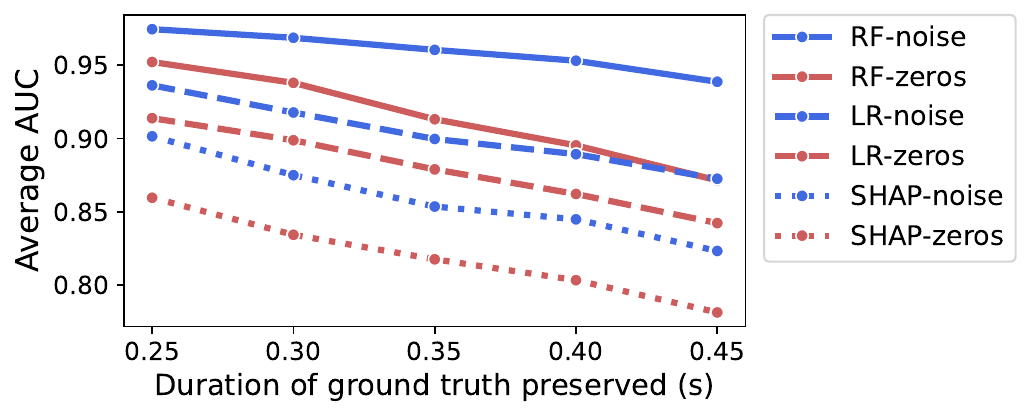}
\vspace{-4mm}
\caption{Average AUC on the Drums dataset as only the first part of the ground truth for each event is preserved.}
\vspace{-6.5mm}
\label{fig:drums_curves}
\end{figure}

\subsection{The impact of inaccurate ground truth} \label{sec:impact-inaccurate-gt}
As mentioned before, the use of AUC over ground truth locations of the events of interest to assess explanation quality implicitly assumes that the full region annotated as containing the event should be important for the model. Yet, for the Drums synthetic dataset we know that only the beginning of the hit, which contains most of the energy, is likely to be relevant for the model. Hence, our ground truth annotations, which are based on the location of the full kick waveform,  include regions that are probably irrelevant. Explainers that find that region to be unimportant would be penalized. 

To analyze the impact of this potential inaccuracy in the ground truth, we computed the average AUC while progressively eliminating the end of each ground truth region from metric computation. In this way, the positive labels are more and more likely to include only the important information for classification. Figure \ref{fig:drums_curves} shows the impact of restricting the labels in this way on the same six explainer models from Figure \ref{fig:auc_relaxed}. We can see that, when keeping only the initial 0.25 seconds of the ground truth, the average AUC of our best model is very close to 1.0. Importantly, the ranking of the systems is preserved even when the ground truth contains non-important segments. This result is essential since it suggests that, as long as the important regions are contained somewhere within the ground truth, the average AUC of the importance values can be used to rank approaches and tune hyperparameters. More evidence is needed to make sure this result holds for other models and datasets.

\subsection{Uncovering spurious correlations}

\begin{figure}[t]
\centering
\includegraphics[width=\columnwidth]{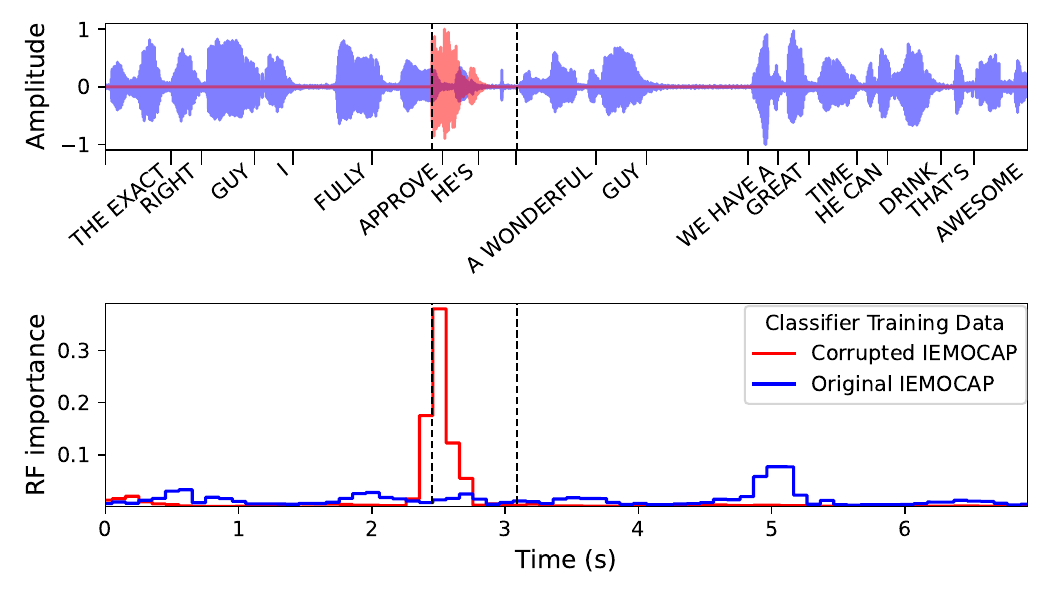}
\vspace{-9mm}
\caption{Example explanation for an IEMOCAP ``happy'' sample with a summed cough (in red) for two models: one trained with the original IEMOCAP samples and one trained with the 
corrupted version that contains coughs for the ``happy'' class.}
\vspace{-7mm}
\label{fig:iemocap}
\end{figure}

Finally, we use the RF+zeros explainer to uncover the spurious correlations in the corrupted IEMOCAP training set (see Section \ref{sec:cleverhans}). Figure \ref{fig:iemocap} shows the importance curves for the emotion classification models trained with the corrupted and the original training sets for the signal shown at the top, where a cough was added to the original IEMOCAP signal. We can see that, for the model trained with the corrupted data, the explanation finds the location of the cough as the most important region, uncovering the fact that the model is relying on this cough to make its decision. Overall, this explainer achieves an average AUC of 0.92 for the detection of the coughs. In a real scenario where this issue was not known to the developer or the user, these explanations would effectively expose the problem.

Notably, the importance curve for the model trained with the original IEMOCAP dataset highlights a part of the waveform that clearly indicates happiness, focusing on the word ``great'' (see the paper's website for the audio waveform) and, to a smaller degree, other three regions over the words ``right'', ``fully approve'' and ``wonderful''.  It also gives some importance to the cough, indicating that the model scores are somewhat affected by it, a reasonable conclusion given that the model has probably seen very few coughs in the training data.

\section{Conclusions}

We proposed a benchmark for assessing the quality of time-localized explanation approaches for audio classification models. We used this benchmark to optimize and compare various explainability approaches by creating a dataset of perturbations of the input sample, using it to train a surrogate model from which explanations are extracted. We found that, in some tasks, the best explanations reach close to perfect performance. Further, we illustrated the use of these models for uncovering spurious correlations in an emotion dataset. The benchmark datasets and code are available on the paper’s website, where we plan to continuously add new datasets and systems.

\section{Acknowledgments}

This material is based upon work supported by the Air Force Office of Scientific Research, United States under the awards number FA9550-18-1-0026 and number F9550-21-1-0445. 

\bibliographystyle{IEEEtran}
\bibliography{mybib}

\begin{thebibliography}{10}
\providecommand{\url}[1]{#1}
\csname url@samestyle\endcsname
\providecommand{\newblock}{\relax}
\providecommand{\bibinfo}[2]{#2}
\providecommand{\BIBentrySTDinterwordspacing}{\spaceskip=0pt\relax}
\providecommand{\BIBentryALTinterwordstretchfactor}{4}
\providecommand{\BIBentryALTinterwordspacing}{\spaceskip=\fontdimen2\font plus
\BIBentryALTinterwordstretchfactor\fontdimen3\font minus \fontdimen4\font\relax}
\providecommand{\BIBforeignlanguage}[2]{{%
\expandafter\ifx\csname l@#1\endcsname\relax
\typeout{** WARNING: IEEEtran.bst: No hyphenation pattern has been}%
\typeout{** loaded for the language `#1'. Using the pattern for}%
\typeout{** the default language instead.}%
\else
\language=\csname l@#1\endcsname
\fi
#2}}
\providecommand{\BIBdecl}{\relax}
\BIBdecl

\bibitem{nasim2024trustworthyxaiapplication}
M.~A.~A. Nasim, P.~Biswas, A.~Rashid, A.~Biswas, and K.~D. Gupta, ``Trustworthy xai and application,'' 2024.

\bibitem{adebayo2020debugging}
J.~Adebayo, M.~Muelly, I.~Liccardi, and B.~Kim, ``Debugging tests for model explanations,'' in \emph{Proceedings of the 34th International Conference on Neural Information Processing Systems}, 2020, pp. 700--712.

\bibitem{islam2022study}
R.~Islam, E.~Abdel-Raheem, and M.~Tarique, ``A study of using cough sounds and deep neural networks for the early detection of covid-19,'' \emph{Biomedical Engineering Advances}, vol.~3, p. 100025, 2022.

\bibitem{Ben-amor2023}
I.~Ben~Amor, J.-F. Bonastre, B.~O’Brien, and P.-M. Bousquet, ``Describing the phonetics in the underlying speech attributes for deep and interpretable speaker recognition,'' in \emph{Interspeech2023}, 2023.

\bibitem{nfissi2023iterative}
A.~Nfissi, W.~Bouachir, N.~Bouguila, and B.~L. Mishara, ``Iterative feature boosting for explainable speech emotion recognition,'' in \emph{2023 International Conference on Machine Learning and Applications (ICMLA)}.\hskip 1em plus 0.5em minus 0.4em\relax IEEE, 2023, pp. 543--549.

\bibitem{surveyxai}
M.~Mersha, K.~Lam, J.~Wood, A.~AlShami, and J.~Kalita, ``Explainable artificial intelligence: A survey of needs, techniques, applications, and future direction,'' 2025.

\bibitem{LRP}
A.~Binder, G.~Montavon, S.~Bach, K.-R. Müller, and W.~Samek, ``Layer-wise relevance propagation for neural networks with local renormalization layers,'' 2016.

\bibitem{deeplift}
A.~Shrikumar, P.~Greenside, and A.~Kundaje, ``Learning important features through propagating activation differences,'' 2019.

\bibitem{IG}
M.~Sundararajan, A.~Taly, and Q.~Yan, ``Axiomatic attribution for deep networks,'' 2017.

\bibitem{ribeiro2016should}
M.~T. Ribeiro, S.~Singh, and C.~Guestrin, ``" why should i trust you?" explaining the predictions of any classifier,'' in \emph{Proceedings of the 22nd ACM SIGKDD international conference on knowledge discovery and data mining}, 2016, pp. 1135--1144.

\bibitem{scott2017unified}
M.~Scott, L.~Su-In \emph{et~al.}, ``A unified approach to interpreting model predictions,'' \emph{Advances in neural information processing systems}, vol.~30, pp. 4765--4774, 2017.

\bibitem{CFE}
S.~Wachter, B.~Mittelstadt, and C.~Russell, ``Counterfactual explanations without opening the black box: Automated decisions and the gdpr,'' 2018.

\bibitem{mishra2017local}
S.~Mishra, B.~L. Sturm, and S.~Dixon, ``Local interpretable model-agnostic explanations for music content analysis.'' in \emph{ISMIR}, vol.~53, 2017, pp. 537--543.

\bibitem{haunschmid2020audiolime}
V.~Haunschmid, E.~Manilow, and G.~Widmer, ``audiolime: Listenable explanations using source separation,'' \emph{CoRR}, vol. abs/2008.00582, 2020.

\bibitem{bento2021timeshap}
J.~Bento, P.~Saleiro, A.~F. Cruz, M.~A. Figueiredo, and P.~Bizarro, ``Timeshap: Explaining recurrent models through sequence perturbations,'' in \emph{Proceedings of the 27th ACM SIGKDD conference on knowledge discovery \& data mining}, 2021, pp. 2565--2573.

\bibitem{chan-etal-2022-comparative}
C.~S. Chan, H.~Kong, and L.~Guanqing, ``A comparative study of faithfulness metrics for model interpretability methods,'' in \emph{Proceedings of the 60th Annual Meeting of the Association for Computational Linguistics (Volume 1: Long Papers)}, S.~Muresan, P.~Nakov, and A.~Villavicencio, Eds.\hskip 1em plus 0.5em minus 0.4em\relax Dublin, Ireland: Association for Computational Linguistics, May 2022, pp. 5029--5038.

\bibitem{paissan2024listenablemapsaudioclassifiers}
F.~Paissan, M.~Ravanelli, and C.~Subakan, ``Listenable maps for audio classifiers,'' 2024.

\bibitem{parekh2022listeninterpretposthocinterpretability}
J.~Parekh, S.~Parekh, P.~Mozharovskyi, F.~d'Alché Buc, and G.~Richard, ``Listen to interpret: Post-hoc interpretability for audio networks with nmf,'' 2022.

\bibitem{Akman2024AttHearEA}
A.~Akman and B.~W. Schuller, ``Atthear: Explaining audio transformers using attention-aware nmf,'' \emph{ICASSP 2024 - 2024 IEEE International Conference on Acoustics, Speech and Signal Processing (ICASSP)}, pp. 7015--7019, 2024.

\bibitem{haufe2024explainableaineedsformal}
S.~Haufe, R.~Wilming, B.~Clark, R.~Zhumagambetov, D.~Panknin, and A.~Boubekki, ``Explainable ai needs formal notions of explanation correctness,'' 2024.

\bibitem{wilming2023theoreticalbehaviorxaimethods}
R.~Wilming, L.~Kieslich, B.~Clark, and S.~Haufe, ``Theoretical behavior of xai methods in the presence of suppressor variables,'' 2023.

\bibitem{abdullah2024sig}
T.~A. Abdullah, M.~S.~M. Zahid, A.~F. Turki, W.~Ali, A.~A. Jiman, M.~J. Abdulaal, N.~M. Sobahi, and E.~T. Attar, ``Sig-lime: a signal-based enhancement of lime explanation technique,'' \emph{IEEE Access}, 2024.

\bibitem{panayotov2015librispeech}
V.~Panayotov, G.~Chen, D.~Povey, and S.~Khudanpur, ``Librispeech: an asr corpus based on public domain audio books,'' in \emph{2015 IEEE international conference on acoustics, speech and signal processing (ICASSP)}.\hskip 1em plus 0.5em minus 0.4em\relax IEEE, 2015, pp. 5206--5210.

\bibitem{lugosch2019speechmodelpretrainingendtoend}
L.~Lugosch, M.~Ravanelli, P.~Ignoto, V.~S. Tomar, and Y.~Bengio, ``Speech model pre-training for end-to-end spoken language understanding,'' 2019.

\bibitem{mcauliffe17_interspeech}
M.~McAuliffe, M.~Socolof, S.~Mihuc, M.~Wagner, and M.~Sonderegger, ``Montreal forced aligner: Trainable text-speech alignment using kaldi,'' in \emph{Interspeech 2017}, 2017, pp. 498--502.

\bibitem{hershey2021benefit}
S.~Hershey, D.~P. Ellis, E.~Fonseca, A.~Jansen, C.~Liu, R.~C. Moore, and M.~Plakal, ``The benefit of temporally-strong labels in audio event classification,'' in \emph{ICASSP 2021-2021 IEEE International Conference on Acoustics, Speech and Signal Processing (ICASSP)}.\hskip 1em plus 0.5em minus 0.4em\relax IEEE, 2021, pp. 366--370.

\bibitem{gong21b_interspeech}
Y.~Gong, Y.-A. Chung, and J.~Glass, ``Ast: Audio spectrogram transformer,'' in \emph{Interspeech 2021}, 2021, pp. 571--575.

\bibitem{drum_dataset}
A.~Chhabra, A.~Veer~Singh, R.~Srivastava, and V.~Mittal, ``Drum instrument classification using machine learning,'' in \emph{2020 2nd International Conference on Advances in Computing, Communication Control and Networking (ICACCCN)}, 2020, pp. 217--221.

\bibitem{liu24f_interspeech}
Y.-L. Liu, R.~Feng, J.-H. Yuan, and Z.-H. Ling, ``Clever hans effect found in automatic detection of alzheimer's disease through speech,'' in \emph{Interspeech 2024}, 2024, pp. 2435--2439.

\bibitem{arjovsky2020invariantriskminimization}
M.~Arjovsky, L.~Bottou, I.~Gulrajani, and D.~Lopez-Paz, ``Invariant risk minimization,'' 2020.

\bibitem{borah2024measuringspuriouscorrelationclassification}
A.~Borah, D.~Pylypenko, C.~Espana-Bonet, and J.~van Genabith, ``Measuring spurious correlation in classification: 'clever hans' in translationese,'' 2024.

\bibitem{WALLIS2022102368}
D.~Wallis and I.~Buvat, ``Clever hans effect found in a widely used brain tumour mri dataset,'' \emph{Medical Image Analysis}, vol.~77, p. 102368, 2022.

\bibitem{gauder2024unreliabilityacousticsystemsalzheimers}
L.~Gauder, P.~Riera, A.~Slachevsky, G.~Forno, A.~M. Garcia, and L.~Ferrer, ``The unreliability of acoustic systems in alzheimer's speech datasets with heterogeneous recording conditions,'' 2024.

\bibitem{busso2008iemocap}
C.~Busso, M.~Bulut, C.-C. Lee, A.~Kazemzadeh, E.~Mower, S.~Kim, J.~N. Chang, S.~Lee, and S.~S. Narayanan, ``Iemocap: Interactive emotional dyadic motion capture database,'' \emph{Language resources and evaluation}, vol.~42, pp. 335--359, 2008.

\bibitem{piczak2015esc}
K.~J. Piczak, ``Esc: Dataset for environmental sound classification,'' in \emph{Proceedings of the 23rd ACM international conference on Multimedia}, 2015, pp. 1015--1018.

\bibitem{ferrer2024goodpracticesevaluationmachine}
L.~Ferrer, O.~Scharenborg, and T.~Bäckström, ``Good practices for evaluation of machine learning systems,'' 2024.

\end{thebibliography}
\end{document}